\documentclass[11pt,twocolumn]{article}

\usepackage{times}

\usepackage{graphicx}

\title{Topological defects produce exotic mechanics in complex metamaterials}

\author
{Anne S. Meeussen,$^{1,2\ast}$ Erdal C.~O\u{g}uz,$^{3}$ Yair Shokef,$^{3}$ Martin van Hecke$^{1,2}$\\
\\
\normalsize{$^{1}$AMOLF, Science Park 104, 1098 XG Amsterdam, the Netherlands,}\\
\normalsize{$^{2}$Huygens-Kamerlingh Onnes Laboratory, Universiteit Leiden,}\\
\normalsize{PO Box 9504, 2300 RA Leiden, the Netherlands}\\
\normalsize{$^{3}$School of Mechanical Engineering and The Sackler Center for Computational 
}\\
\normalsize{Molecular and Materials Science,Tel Aviv University, Tel Aviv 6997801, Israel}\\
\\
\normalsize{$^\ast$Correspondence should be addressed to:  a.meeussen@amolf.nl.}
}

\date{}

\begin{document}

\maketitle 

\renewcommand{\figurename}{\textbf{Fig.}}
\renewcommand\thefigure{\textbf{\arabic{figure}}}


\textbf{
  Defects, and in particular topological defects, are architectural motifs that play a crucial role in natural materials.  Here we provide a systematic strategy to introduce such defects in mechanical metamaterials. We first present metamaterials that are a mechanical analogue of spin systems with tunable ferromagnetic and antiferromagnetic interactions,	then design an exponential number of frustration-free metamaterials, and finally introduce topological defects by rotating a string of building blocks in these metamaterials. We uncover the distinct mechanical signature of topological defects by experiments and simulations, and leverage this to design complex metamaterials in which we can steer deformations and stresses towards parts of the system. Our work presents a new avenue to systematically include spatial complexity, frustration, and topology in mechanical metamaterials.
}

Mechanical metamaterials are structured forms of matter with unprecedented properties, including negative response parameters~\cite{Mullin2007,Shim2012}, shape-morphing~\cite{Coulais2016a,Dudte2016c}, topological mechanics~\cite{Chen2016,Paulose2015a,Paulose2015,Serra-Garcia2018}, self-folding~\cite{Coulais2018c}, and programmability~\cite{Florijn2014a,Silverberg2014,Frenzel2017,Bertoldi2017}. So far, the focus has been on compatible architectures, where all the unit cells can deform in harmony. However, frustration can yield advanced functionalities, such as multistability and programmability \cite{Florijn2014a,Silverberg2014,Kang2014}, and may open up opportunities to probe controlled frustration in man-made systems~\cite{Nisoli2013,Wang2006,Castelnovo2008}. We note that deformations of unit cells in compatible metamaterials often alternate, leading to horizontal and vertical ellipses~\cite{Mullin2007,Shim2012}, rigid elements that rotate left or right~\cite{Frenzel2017,Grima2005,Coulais2018a}, or edges of unit cells that move in or out~\cite{Coulais2016a}: we refer to this as \emph{antiferromagnetic order}. Hence, a promising route to introduce targeted frustration is to introduce unit cells that can favor either ferromagnetic, non-alternating order or antiferromagnetic order,
and use the freedom supplied by additive manufacturing to stack these at will.

To implement this strategy we employ anisotropic, triangular building blocks that set the ferromagnetic or antiferromagnetic nature of their interactions depending on their mutual orientations. These building blocks consist of six elastic \emph{edge bonds} and two elastic \emph{internal bonds} connected by flexible hinges, or \emph{nodes}---each building block has one \emph{minority} node that determines its
orientation (Fig.~1A). Each building block features a floppy ``hinging'' mode, which allows for finite deformations while bond lengths remain constant. Accordingly deformed building blocks can take on two shapes---fat and skinny---which we assign a positive or negative spin variable (Fig.~1B). Two adjacent building blocks exhibit a collective zero-energy deformation. When their shared edge contains one minority node, the building blocks interact ferromagnetically and the zero-energy deformation features two building blocks with the same spin; in contrast, when their shared edge possesses zero or two minority nodes, the interaction is antiferromagnetic and the deformations have opposite spins (Fig.~1C).

We generate a family of metamaterials by combining building blocks to create mechanical networks that realize complex spin systems on a hexagonal lattice. Random configurations are typically highly frustrated. To obtain targeted frustration, we first design frustration-free configurations by choosing the building-block orientations so that the deformed building blocks fit as in a jigsaw puzzle. One example of a frustration-free geometry is an ordered metamaterial where all interactions are antiferromagnetic (Fig.~1D).
Its zero-energy deformation mode has all up-facing (down-facing) building blocks in their fat (skinny) state, or vice-versa, which corresponds to alternating positive and negative spins (Fig.~1D). We note that this geometry is equivalent to the rotating square mechanism that underlies the design of a wide range of metamaterials \cite{Mullin2007,Bertoldi2010,Coulais2018c,Florijn2014a,Shim2012,Grima2005,Coulais2018a}.

We can design a vast number of structurally complex yet frustration-free configurations by requiring that in each hexagon of six adjacent building blocks, the number of antiferromagnetic interactions is even. We note that in all configurations the edge bonds form a triangular scaffold, while the internal bonds within each hexagon form closed \emph{local loops} around each vertex of the scaffold (Fig.~1, D and E). The direct relation between the sign of the interaction, the shared minority nodes, and the internal bonds implies that compatibility is equivalent to requiring that all local loops are even (Fig.~1, D and E; see supplementary materials). We 3D printed one such compatible metamaterial, realizing bonds by elastic beams and hinges by thin joints, and find that external forcing excites the designed soft mode  (Fig.~1,E and F; see supplementary materials).

Designing and counting complex frustration-free geometries requires solving combinatorial problems. For example, geometries with
fully antiferromagnetic spins can be mapped to diamond tilings, where each diamond represents two building blocks with two minority nodes on their shared edge (see supplementary materials). Moreover, as all compatible architectures feature a floppy mode where all building blocks have two edge-nodes moving ``in'', and one edge-node moving ``out'', or vice versa, we can map these deformations to ground states of the antiferromagnetic Ising model on the kagome lattice (Fig.~1G),
so that each ground state generates a distinct compatible metamaterial (see supplementary materials).
The extensive ground-state entropy of that model~\cite{Syo51,Kan53} yields an asymptotically exact result for the exponential number of compatible architectures as a function of the system size~\cite{Riv16} (Fig.~1H).

Our combinatorial design allows us to introduce frustration in a targeted manner by discrete rotations of the building blocks of compatible architectures, irrespective of their spatial complexity. We first note that in a frustration-free design, all even-length local loops are connected, such that larger paths enclosing multiple local loops are also even (Fig.~2A). Rotating a single building block generates a \emph{structural defect} associated with two adjacent odd local loops (Fig.~2B). These odd loops eliminate the floppy mode and  lead to a stiffening of the metamaterial. However, larger loops around such a structural defect are still even, indicating only a local breakdown of compatibility. Rotating consecutive blocks can drive one of the odd loops to the boundary of the system, and the resulting metamaterial is left with a single local loop of odd length (Fig.~2C). We interpret this as a \emph{topological defect}~\cite{Mermin1979,Alexander2012}, which affects the metamaterial at the global scale, since all loops of internal bonds surrounding the defect have odd perimeter. A topological defect thus results in long-ranged conflicts that cannot be resolved through local design changes, such as rotation of a few building blocks. We note that, in contrast to defects occurring in metamaterials where the nontrivial topology results from a nonzero winding number in momentum space~\cite{Paulose2015a,Paulose2015,Kane2013}, here the topological character of defects is governed by the parity of real-space local loops.

Metamaterials with structural, topological, or no defects have distinct mechanical signatures. We detect these by considering pairs of building blocks at the system's boundary, deforming one building block and tracking the deformation of the other. Specifically, we extend a block $i$ by forcing its majority nodes from length $l$ to $l(1+\delta_i)$, measure the resulting deformation $\delta_j$ of block $j$, and define a deformation transfer factor $q_{ij}= \delta_j / \delta_i$  (Fig.~3A). In a compatible metamaterial of freely hinging springs, $q_{ij}=\pm 1$, while incompatibilities or bending interactions cause $|q_{ij}|<1$. Crucially, the sign of $q_{ij}$ reflects the nature of the interactions between \emph{neighboring} blocks $i$ and $j$, being antiferromagnetic (ferromagnetic) if $q_{ij}<0$ ($q_{ij}>0$). We separately measure the transfer factors between all neighboring pairs of the $P$ blocks around the network's perimeter; define a cumulative product $q_n:=\prod_{i=1}^{n} q_{i,i+1}$ that relates block 1 to block $n+1$; and introduce
a normalized order parameter ${Q} := sign(q_P) \cdot |q_P|^{1/P}$ that characterizes the full perimeter.

The magnitude and sign of ${Q}$ distinguish metamaterials with structural, topological, or no defects. We first consider idealized metamaterials consisting of Hookean springs connected by Freely Hinging nodes (model FH, see supplementary materials). The sign of $q_n$ can vary wildly with $n<P$, reflecting the mixed antiferromagnetic and ferromagnetic interactions in our designs (Fig.~3B). However, the sign of $Q$ precisely measures the parity of the closed loop of internal bonds around the boundary, being positive for a structural defect and negative for a topological defect.
For a compatible system, deformations follow the global floppy mode, all building blocks deform with equal magnitude and, since any loop around the system boundary is even, $Q=1$.
An incompatible network has no global floppy mode, hence deformations decay away from the actuation point so that $|q_{ij}|<1$ and $|{Q}| < 1$; crucially, the sign of $Q$ is not sensitive to this decay. Structural and topological defects can thus be distinguished by the sign of ${Q}$ (Fig.~3C).

In experiments, $|q_{ij}|<1$ due to elastic decay that stems from finite Torsional Resistance of the hinges~\cite{Coulais2018a}. Numerical calculations on mechanical networks that include these additional interactions (model TR, see supplementary materials) reproduce this effect. The elastic decay blurs the distinction between defect-free metamaterials and those with a single structural defect, as both have $0<Q<1$. Nevertheless, our method allows us to unambiguously detect topological defects, for which ${Q}<0$ (Fig.~3, B and C). We can also detect topological defects by considering significantly fewer than all $P$ edge blocks, making our detection scheme practical and robust (see supplementary materials). Thus, the nontrivial frustration associated with topological defects can be successfully distinguished by mechanically probing the metamaterial's edge.

Finally, we design metamaterials in which we can steer deformations and stresses by extending or contracting a small number of building blocks. Consider a metamaterial that connects two triangular building blocks $i,j$ via two compatible strips. If the interactions between $i$ and $j$ are ferromagnetic in one strip and antiferromagnetic in the other, this geometry contains a topological defect in the excised center; while if the interactions are both (anti)ferromagnetic, the geometry is compatible and topologically trivial (Fig.~4, A and B). In the latter, compatible case, simulations using model FH (Fig.~4A), model TR, and experiments (see supplementary materials) show that simultaneously forcing blocks $i$ and $j$ will either lead to large deformations, or large stresses, depending on the parity of the forcing. In the former, topologically nontrivial case, forcing with the same parity will lead to large deformations in one strip, and large stresses in the other---changing the parity of forcing flips the dominance of deformations and stresses (Fig.~4B; see supplementary materials). Geometries with more defects, holes and actuation sites can be designed to arbitrarily steer stresses and displacements across larger systems.

Our work shows how sufficiently complex unit cells can be combined into metamaterials with targeted, discretely controlled frustration and nontrivial topology. This strategy allows for designing novel classes of frustrated metamaterials, including origami, kirigami and 3D metamaterials \cite{Coulais2016a,Bertoldi2017,Ning2018} for future technologies concerning sensing, actuation, and soft robotics \cite{McEvoy2015,Reis2015,Wehner2016}. More generally, our strategy opens up a new avenue for studying topological, spatially complex states in artificial materials that are experimentally accessible.

\begin{figure*}
	\centering	
	\includegraphics[]{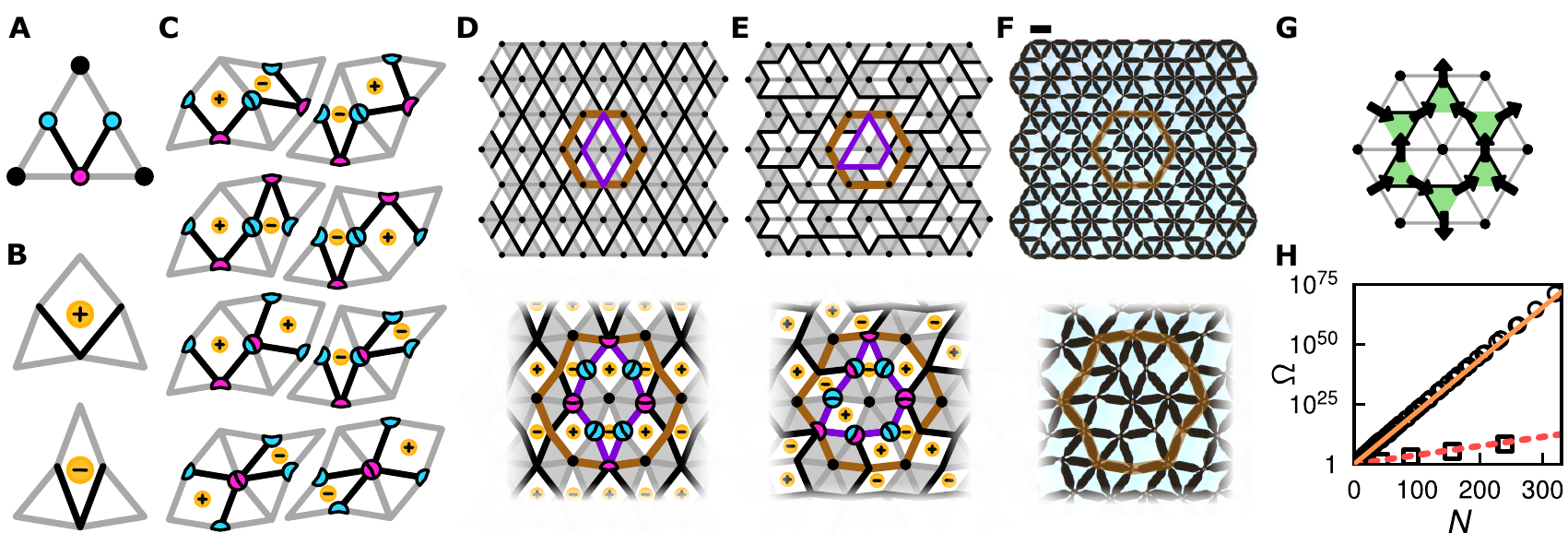}
	\caption{\baselineskip24pt {\bfseries Structurally complex compatible metamaterials.} (\textbf{A}) Anisotropic building blocks consisting of edge bonds (grey), internal bonds (black), corner nodes (black), majority edge-nodes (blue), and a minority edge-node (pink). (\textbf{B}) Deformed building block in fat (+) and skinny (-) states. (\textbf{C}) Adjacent building blocks interact antiferromagnetically (ferromagnetically) when their shared edge features an even (odd) number of minority nodes, so that their minority nodes are connected by an even (odd) number of internal bonds. (\textbf{D}) Ordered configuration, with local loop (purple), associated hexagon (brown) and rigid triangles (grey). Zoom-in: antiferromagnetic ordering and concomitant real-space deformation. (\textbf{E}) Structurally complex architecture. Zoom-in: mixed ferromagnetic and antiferromagnetic ordering and deformation. (\textbf{F}) Experimental realization of the compatible metamaterial in panel (\textbf{E}). Scale bar: 1 cm. Zoom-in: deformations due to uniform compression of the left and right edges (see supplementary materials) show close agreement between experimental observation and theoretical prediction. (\textbf{G}) Node deformations can be mapped to Ising spins (arrows) on the kagome lattice (green triangles). (\textbf{H}) Numerically obtained number of compatible designs $\Omega$ as a function of the number of edge-nodes $N$ (circles), compared to asymptotic prediction (orange line)~\cite{Syo51}, based on the Ising-spin mapping; and exact (squares) and asymptotic (red dashes) number of fully antiferromagnetic designs (see supplementary materials).
	}
\end{figure*}

\begin{figure}
	\centering
	\includegraphics[]{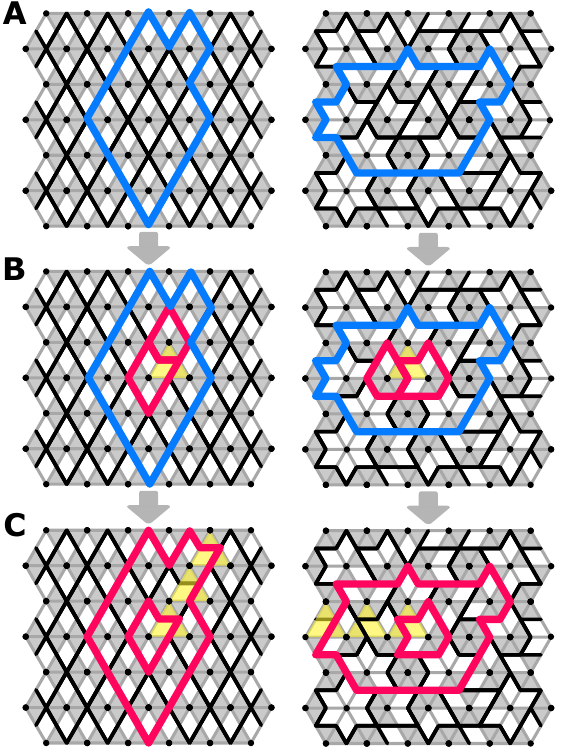}
	\caption{\baselineskip24pt {\bfseries Defect design.} (\textbf{A}) Two compatible network designs where all loops, such as the highlighted blue one, are of even length. (\textbf{B}) Rotating a single building block (yellow) gives rise to a structural defect associated with two adjacent local odd loops (red). However, larger loops (blue) encircling these local loops remain even. (\textbf{C})  Rotating consecutive blocks (yellow) can produce a topological defect characterized by a single local odd loop. Any loop encircling the topological defect is odd (red).
	}
	\label{fig3}
\end{figure}

\begin{figure*}
	\centering
	\includegraphics[]{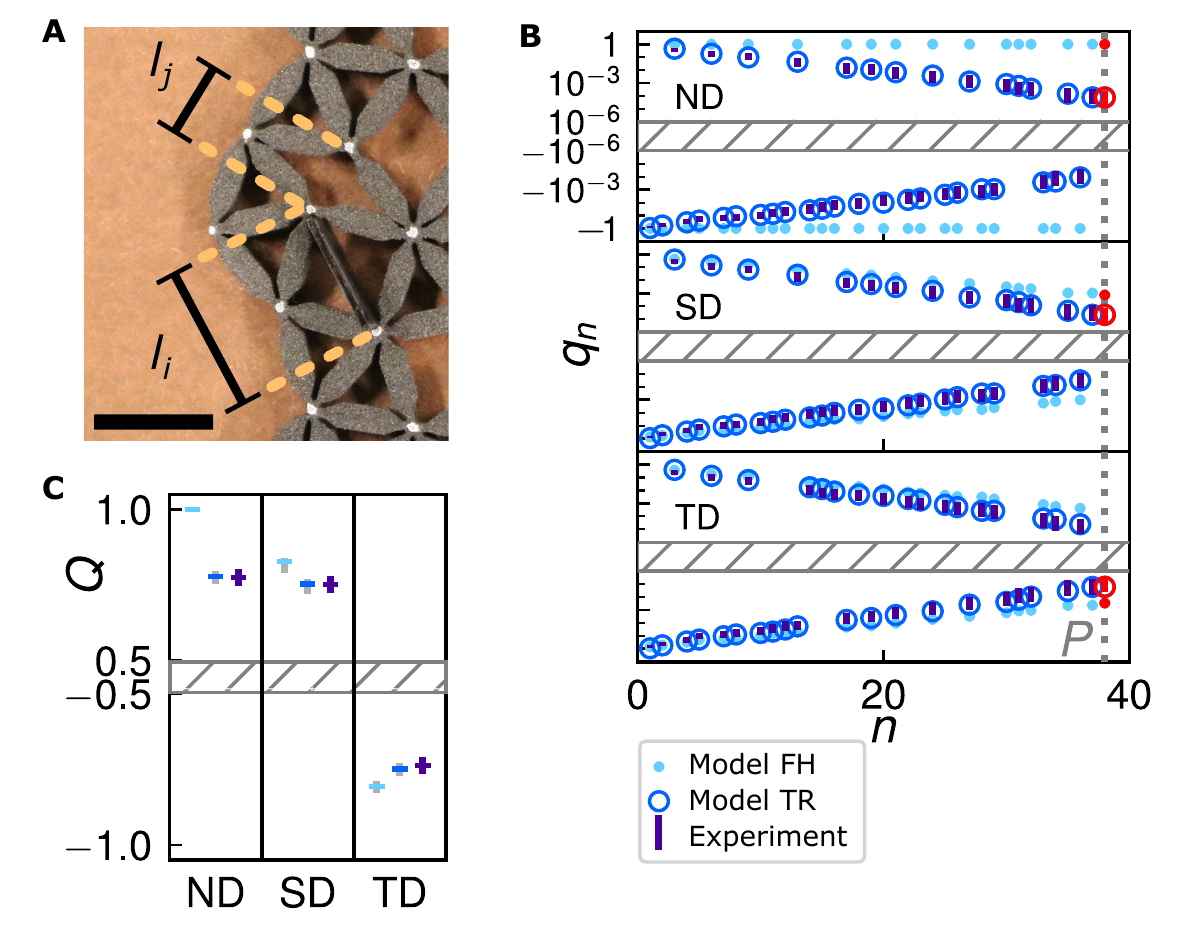}
	\caption{\baselineskip24pt {\bfseries  Defect detection.} (\textbf{A}) Zoom-in of metamaterial, showing two edge-nodes forced apart (by an inserted wedge) over a distance $l_i$. The response is characterized by $l_j$. Scale bar: 1 cm. (\textbf{B}) Cumulative product $q_n$ (on log scales) as function of probed edge blocks $n$ for three architectures: with no defect (ND), a structural defect (SD) and a topological defect (TD). In all cases, we compare experiments (bars proportional to standard deviaton) to simulations of spring-networks with and without torsional rigidity (model TR and model FH, respectively). The data shows a striking coincidence of the signs of $q_n$, even though their decay with $n$ depends on details. The sign of $q_P$ (red) clearly distinguishes between topologically non-trivial and trivial states. (\textbf{C}) Order parameter $Q$ for the three architectures (colors: see legend). Markers (grey) on model data show the spread of $Q$ calculated for 99 distinct networks of each type; error bars on experimental data show standard deviaton. 
		\label{fig4}}
\end{figure*}

\begin{figure*}
	\centering
	\includegraphics[]{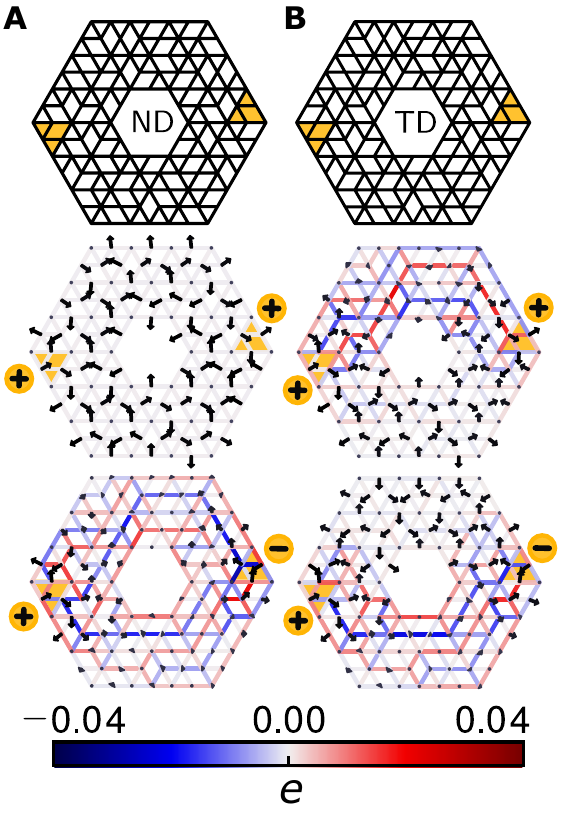}
	\caption{\baselineskip24pt {\bfseries  Stiffness steering.} (\textbf{A}) Compatible architecture with center removed (ND, top) shows a homogeneous displacement (arrows) and absence of bond stretching $e$ (colors) for compatible forcing (middle) at two blocks (yellow triangles) according to model FH, and significant bond stretching for incompatible forcing (bottom). (\textbf{B}) Topologically non-trivial architecture where a topological defect was placed in the removed center (TD, top). Under simultaneous extension of the two blocks (yellow markers), bond stretching localizes in the upper portion of the sample, while node deformations localize in the lower part (middle); under actuation with opposite parity, stretching localizes in the lower and deformations in the upper part (bottom).
		\label{fig4}}
\end{figure*}

\clearpage

\renewcommand{\refname}{References and notes}

\clearpage 

\section*{Acknowledgments}
We thank Roni Ilan, Edan Lerner, Bela Mulder, Ben Pisanty, Eial Teomy and Ewold Verhagen for fruitful discussions, Jayson Paulose for co-developing code for the numerical model FH, Dion Ursem for technical support, and Rivka Zandbergen for supplying the exact counting data in Fig.~1H. This research was supported in part by the Israel Science Foundation Grant No. 968/16. The authors declare no competing interests. Requests for materials should be addressed to A.S.M. (a.meeussen@amolf.nl).

\setcounter{figure}{0}
\renewcommand{\figurename}{\textbf{Figure}}
\renewcommand\thefigure{\textbf{S\arabic{figure}}}
\setcounter{table}{0}
\renewcommand{\tablename}{Table}
\renewcommand\thetable{S\arabic{table}}

\section*{Supplementary Information}

\subsection*{Parity of paths of internal bonds}
In a compatible metamaterial, all building blocks can simultaneously deform according to their local floppy mode. We conceive of the joint floppy deformation of any pair $i,j$ of blocks as an interaction: for a \emph{ferromagnetic} interaction, the blocks simultaneously expand or contract, whereas for an \emph{antiferromagnetic} interaction one block contracts while the other expands.

The internal bonds connecting blocks $i,j$ determine their interaction type: for a single building block, two edges connected by a single bond display an opposite ``in'' and ``out'' motion, whereas two edges connected via two internal bonds display the same ``in'' or ``out'' motion. Consider a path of $N_b$ internal bonds running from the minority node  of  block $i$ to the minority node of block $j$, and define $N_T$ as the number of triangular building blocks (including block $i$ and $j$). The  \emph{path parity} $\Pi= (-1)^{N_b-N_T-1}$ is then positive (negative) when the interaction between $i$ and $j$ is ferromagnetic (antiferromagnetic), see Extended Data Fig.~S1, A and B.

Compatibility requires that, along a closed local loop within a hexagon of six adjacent blocks, each block interacts ferromagnetically with itself. Equivalently, the parity of internal bonds in the loop is even, as is the number of minority nodes on shared edges, as well as the number of (anti)ferromagnetic interactions between adjacent blocks.

\subsection*{Experimental realizations}
Experimental realizations of our complex metamaterials are produced using a Sinterit Lisa 3D printer with thermoplastic polyurethane powder Sinterit Flexa Black, processed at a sintering layer height of $0.1~\mathsf{mm}$ at the ``softer'' setting~\cite{Sinterit,Sinterit_flexa}. The reported Young's modulus of the base material printed at these specifications is $4 \pm 0.5~\mathsf{MPa}$, and the final printed networks have a height of $5\pm0.2~\mathsf{mm}$. The individual bars of the network are realized as thick beams, connected by thin beams at each joining node, marked in white (Extended Data Fig.~S2, A and B).

In homogeneous compression experiments, the network is placed between two parallel rigid blocks on its left and right sides. The sample shown in Fig.~1F (lateral dimension  121 mm) is compressed to 118 mm. The sample is imaged by a Canon EOS 750D camera, taking $6000$ by $4000~\mathsf{px}$ images before and after actuation at a resolution of $33~\mathsf{px/mm}$. When actuating individual building blocks, the node positions are extracted from both images and used to calculate a displacement vector associated with each node, after subtracting rigid-body translations and rotations. From this data we extract the initial and final lengths $l_{i,0}$ and $l_i$ of the distances between the internal nodes of each building block to determine the block strains $\delta_i = l_i/l_{i,0}-1$.

For the topological detection scheme, the network is actuated by consecutively inserting a $12.4\pm0.2~\mathsf{mm}$-wide stiff wedge into each edge block, separating its internal nodes from $10 \pm0.2~\mathsf{mm}$ to $13.4 \pm0.4~\mathsf{mm}$. The node positions and edge block strains are extracted to determine the transfer factors $q_{ij} = \delta_j/\delta_i$. For the stress-steering scheme, two blocks are simultaneously actuated by inserting wedges of $12.5\pm0.2~\mathsf{mm}$ width, inducing an internal node separation from $10 \pm0.2~\mathsf{mm}$ to $13.5 \pm0.4~\mathsf{mm}$.

The experimental measurements of the initial internal node spacings $l_{i,0}$ are normally distributed with a standard deviation of $\Delta l=0.15~\mathsf{mm}$. From this we estimate errors for the experimentally measured transfer factors $q_{ij}$, cumulative transfer product $q_n$, and order parameter $Q$ as $ \Delta q_{ij}^2 \approx \Delta l^2 q_{ij}^2 A_{ij} : A_{ij} = \left( \frac{(l_j/l_{j,0})^2+1}{(l_j - l_{j,0})^2} + \frac{(l_i/l_{i,0})^2+1}{(l_i - l_{i,0})^2}\right)$, $\Delta\log(q_n)^2 \approx \Delta l^2\sum_{ij}^{n} A_{ij}$ and $\Delta Q^2 \approx Q^2 \frac{\Delta l^2}{n^2}\sum_{ij}^{n} A_{ij}$ respectively.

\subsection*{Compatible metamaterials with fully antiferromagnetic interactions}
We map the design of antiferromagnetic architectures, where all neighboring building blocks interact antiferromagnetically, to the tiling of diamonds. Since only building blocks that share zero or two minority nodes interact antiferromagnetically, each building block needs to be oriented so that its minority node is paired with the minority node of one of its neighbors. Identifying such pairs of building blocks as a diamond-shaped tile (Extended Data Fig.~S3A), each antiferromagnetic architecture maps to a unique tiling of diamonds. Counting the number of antiferromagnetic architectures thus corresponds to counting diamond tilings, a partition problem of considerable interest in statistical and condensed matter physics~\cite{Blunt1077}.

Solutions to this problem yield the number of antiferromagnetic architectures, $\Omega_{AF}$, as a function of the number of edge nodes $N$. The number of \emph{hexagonal} diamond tilings $\Omega_{AF}$ with $n$ diamonds along each hexagon side (e.g. Extended Data Fig.~S3B for $n=2$) can be calculated exactly~\cite{macmahon} to be
\begin{equation}
\Omega_{AF} =  2 \prod_{i=1}^{n}\prod_{j=1}^{n}\prod_{k=1}^{n} \frac{i+j+k-1}{i+j+k-2},
\end{equation}
which approaches an exponential function in the thermodynamic limit~\cite{oeis}  (Fig.~1H):
\begin{equation}
\Omega_{AF} \sim e^{\log(3^{1/2}/2^{2/3})N} \approx e^{0.087N},
\end{equation}
where $N = 3n(3n+1)$.

\subsection*{Mapping compatible metamaterials to an antiferromagnetic Ising model on the kagome lattice}
Designing compatible architectures corresponds to finding ground states of the antiferromagnetic Ising spin model on the kagome lattice (AFIK model). We associate a positive (negative) binary edge spin variable to an outward (inward) deflection in a down-pointing building block, and vice versa for up-pointing blocks (Extended Data Fig.~S4A); each pair of opposite edge spins is connected by an internal bond. Adjacent blocks deform compatibly when their shared edge spins match. Thus, in compatible architectures, the edge spins form a kagome lattice where each triangular plaquette features one ``in'' and two ``out'' edge spins, or vice versa (Extended Data Fig.~S4B). Such edge spin states are precisely the degenerate ground states of the AFIK model, so that each ground-state configuration generates a distinct compatible metamaterial (up to a global spin flip).

From the AFIK mapping, we obtain an asymptotic expression for $\Omega_0$, the number of compatible architectures, via the residual entropy $S_0 \approx 0.502N$ of the degenerate ground state of the AFIK model~\textit{(21,22)}
\begin{equation}\label{eq2}
\Omega_0 \sim e^{0.502N} = e^{0.753T} \approx 2.1^T  < 3^T = \Omega_{tot},
\end{equation}
where $N$ denotes the number of edge spins, $T$ the number of blocks, $N=3T/2$ the number of edge spins in the thermodynamic limit, and $\Omega_{tot}$ the total number of architectures. The asymptotic expression agrees well with the exact number of compatible, parallelogram-shaped architectures as determined by sophisticated computer algorithms~\textit{(23)}, even for small systems (Fig.~1H).

While $\Omega_{AF}$ counts hexagonal systems and $\Omega_0$ parallelogram systems, both expressions grow exponentially with system size; boundary effects are expected to be negligible, and this shape distinction should vanish for large $N$. We find that $\Omega_0 \ll \Omega_{AF}$, so that in the thermodynamic limit a vanishing fraction of all compatible architectures has a purely antiferromagnetic interaction pattern.

\subsection*{Model FH: spring network simulations}
In \emph{model FH}, we capture the linear response of our systems based on networks of Hookean springs connected by freely hinging nodes~\cite{Pellegrino1993}, such that each bond contributes a potential stretching energy $\epsilon_s = \frac{k_s}{2} e^2 $, where $k_s$ is the bond's stiffness and $e$ the elongation from its equilibrium length.

Each node $i$ supplies two degrees of freedom via spatial displacements $u_{x,i}$ and $u_{y,i}$, while a bond between two nodes at locations $\mathbf{r}_i$ and $\mathbf{r}_j$ constrains these motions by resisting linearized bond elongation $e_{ij} = \frac{\mathbf{r}_i-\mathbf{r}_j}{|\mathbf{r}_i-\mathbf{r}_j|}\cdot(\mathbf{u}_i - \mathbf{u}_j)$. For a large network, the vector of bond elongations $\mathbf{e}=(...e_{ij}...)$ is related to the vector of nodal displacements $\mathbf{u}=(...u_{x,i},u_{y,i}...u_{x,j},u_{y,j}...)$ via a compatibility matrix $\mathbf{R}$ so that $\mathbf{e} = \mathbf{R} \mathbf{u}$. Bond elongations result in bond tensions $\tau_{ij}$ via a constitutive equation: $\mathbf{\tau} = \mathbf{K} \mathbf{e}$, where $\mathbf{K}$ is a diagonal matrix of bond stiffnesses that we set equal to the identity. Tensions are in turn converted to nodal loads $(f_{x,i},f_{y,i})$ via $\mathbf{f} = \mathbf{R}^{T}\mathbf{\tau}$.

We use the following approach to calculate the network's behavior when selected nodes are displaced, while the rest are free to move. First, we calculate node forces $\mathbf{f}$ resulting from imposing an initial displacement $\mathbf{u} = \sum_{i}\mathbf{u}_i$, where $\mathbf{u}_i = (0...u_i^x, u_i^y,...0)$ are desired individual node displacements. Second, we determine how the network relaxes to mechanical equilibrium so that node forces vanish except along the forcing directions: an appropriate compensating force is obtained via $\mathbf{f}_{p} = \mathbf{f} - \mathbf{N}_{p} \mathbf{N}_p^T\mathbf{f}$, where $\mathbf{N}_p$ is the matrix with forcing directions $\mathbf{\hat{n}}_i = (0...n_i^x, n_i^y,...0)$ as its columns. The corresponding relaxation displacement $\mathbf{u}_p$ is then calculated from the compensating force $\mathbf{f}_p$ and the reduced compatibility matrix $\mathbf{R}_{p}= \mathbf{R} - \mathbf{R}\mathbf{N}_{p} \mathbf{N}_p^T $. Lastly, the final displacement state of the network is given by $\mathbf{u}_{\mathrm{full}} = \mathbf{u} + \mathbf{u}_{p}$, and the matching bond elongations, tensions and node forces can be obtained from this displacement state. The resulting network response, valid in the regime of small deformations, is compatible with both the imposed node displacement and the conditions of mechanical equilibrium.

\subsection*{Model TR: spring network simulations with torsional hinge rigidity}
In our experimental metamaterials, hinges cost energy to deform. We capture this cost in \emph{model TR}, in which bonds are modelled by Hookean springs as in model FH, but an energy contribution $\epsilon_h= \frac{k_h}{2}\Delta\phi^2$ is added. Here, $k_h$ is the torsional hinge rigidity and $\Delta\phi$ is the deviation of the angle between two neighboring bonds from its equilibrium value. Hence, the total potential energy of a modelled network is $\epsilon = \sum_b \epsilon_s + \sum_{\alpha} \epsilon_h$, where the first sum runs over all bonds, and the second sum over all angles between neighboring pairs of bonds.

To obtain a network's configuration under actuation, we use a standard simulated annealing algorithm, to minimize its total potential energy by probabilistically updating the spatial coordinates of a randomly chosen node at each step, with Metropolis dynamics and a dimensionless pseudotemperature decreasing gradually to zero over $50 \cdot 10^6$ steps.

The modelled response crucially depends on the dimensionless stiffness ratio $\tilde{k} = k_sl_0^2/k_h$. To accurately model our experimental findings, we estimate the order of magnitude of $\tilde{k}$ in our 3D printed networks by assuming all torsional and stretching deformations take place in the hinges, which have thickness $t$, length $l$, and width $w$ and are made of a material with Poisson's ratio $\nu$ and Young's modulus $E$. To linear order, the bending and stretching stiffnesses of such a hinge are given by~\cite{Audoly2010_bend} $k_h = E t^3 w/[12 (1-\nu^2) l]$ and $k_s = E t w/l$, resulting in a stiffness ratio $\tilde{k}=12(1-\nu^2)l_0^2/t^2$. Using the experimental values $l_0 = 10 \pm 0.2~\mathsf{mm}$,  $t = 0.7 \pm 0.2~\mathsf{mm}$, and an experimentally estimated Poisson's ratio of $\nu=0.42\pm0.02$, we estimate $\tilde{k} \approx 2000$. To refine this estimate, we compute the order parameter $Q$ at various values of $\tilde{k}$ for networks corresponding to the experimental sample designs and compare the resulting values of $Q$ to their experimental counterparts. The best match is found at $\tilde{k} \approx 3200$, of the same order of magnitude as the initial estimate (Extended Data Fig.~S5), and this stiffness ratio is subsequently used for model TR.

\subsection*{Topological detection with fewer blocks}
The mechanical detection scheme (main text) to distinguish networks with compatible or defected architectures can be successfully performed by probing only a few of the network's $P$ boundary building blocks. We choose a subset of $B$ roughly equally spaced boundary blocks, actuating each block by extending it to a block strain of $\delta_i$, and calculate the transfer factors $q_{ij} = \delta_j/\delta_i$ for each block pair. We define a diluted cumulative transfer product $q^B_n:=\prod_{i,j}^{n} q_{ij}$ and normalized order parameter $Q(B) = sign(q^B_B) \cdot |q_B^B|^{1/P}$, that are expected to show the same behavior as the full cumulative transfer product $q_n$ and order parameter $Q$ when all edge blocks are probed.

To demonstrate the effectiveness of our detection scheme when only $B$ of all $P$ network boundary blocks are probed, numerical models FH and TR as well as experiments were used to calculate $q^B_n$ and $Q(B)$ for various fractions of $B/P$ (Extended Data Fig.~S6, A and B). The data indicate that model TR shows excellent quantitative agreement with the experimental results, while model FH agrees qualitatively well. Moreover, the cumulative transfer product over $n$ adjacent edge blocks is the same as that for two edge blocks separated by a distance of $n$ blocks, implying an exponential decay in edge block strains away from the actuation point with a characteristic length proportional to $-1/\ln{|Q|}$. Lastly, the diluted order parameter $Q(B)$ has a sign that is independent, and a magnitude that is nearly independent, of the number of probed edge blocks, indicating that $Q(B)$ is the correct normalized order parameter for our systems.

\subsection*{Stress and deformation steering}
Designing path parity between two block actuation sites allows us to control where stresses and deformations localize. We control path parity by either switching the forcing direction of one of the two blocks (main text), or by selecting a new, suitable block pair while fixing the forcing direction. The latter is experimentally more convenient, and produces the same steering effects, as we show here. Two suitable pairs of building blocks are selected in a network with no defect (Extended Data Fig.~S7A). The blocks are part of a closed loop of an even number of internal bonds, which is separable into two connecting paths along the top and bottom of the sample. Both paths must have the same positive or negative parity, corresponding to a ferromagnetic or antiferromagnetic interaction between the two blocks. When a topological defect is present (Extended Data Fig.~S7B), the blocks are connected in a closed loop of odd length so that the two paths have opposite parity: one corresponds to an antiferromagnetic block interaction while the other is ferromagnetic, or vice versa. When a block pair is forced ferromagnetically by expanding both, the ferromagnetic connecting paths are compatible with the forcing and deform easily, while the antiferromagnetic paths are highly frustrated and hardly deform. The networks' response to ferromagnetic block forcing in experimental samples, model TR, and model FH demonstrate a good agreement with the above predictions and confirm the validity of our design approach (Extended Data Fig.~S7, C to H).

\clearpage

\section*{Supplementary Figs. S1-S7}

\begin{figure}[htb!]
	\centering
	\includegraphics[width=\linewidth]{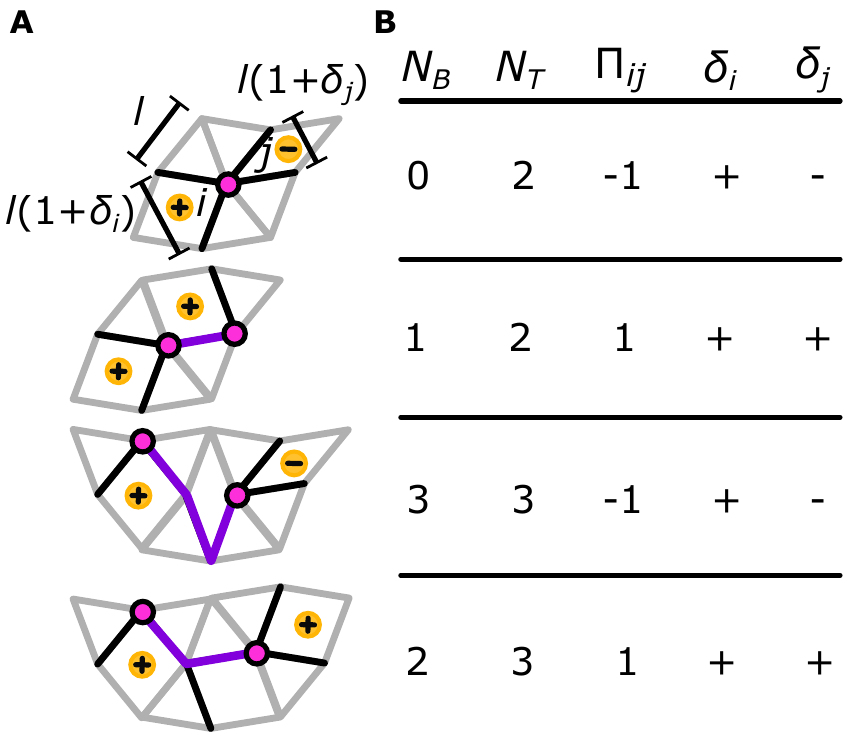}
	\caption{ {\bfseries Path parity.} (\textbf{A}) Four compatible networks are shown. When connected blocks $i,j$ undergo a floppy deformation, their internal nodes deform from initial separation $l$ to $l(1+\delta_i)$ and $l(1+\delta_j)$, where $\delta_i>0$ ($<0$) signifies extension (contraction). (\textbf{B}) The relative sign of $\delta_i$ and $\delta_j$ depends on the length $N_B$ of the connecting path of internal bonds (purple), which runs from the minority node of block $i$ to that of block $j$ (pink), and the number of connecting blocks $N_T$. A path parity $\Pi= (-1)^{N_b-N_T-1}$ may be defined so that the signs of $\delta_i, \delta_j$ are identical (different) if $\Pi=1$ ($\Pi=-1$), corresponding to a ferromagnetic (antiferromagnetic) interaction between $i$ and $j$.}
	\label{fig_meth_A02}
\end{figure}

\begin{figure}[htb!]
	\centering
	\includegraphics[width=\linewidth]{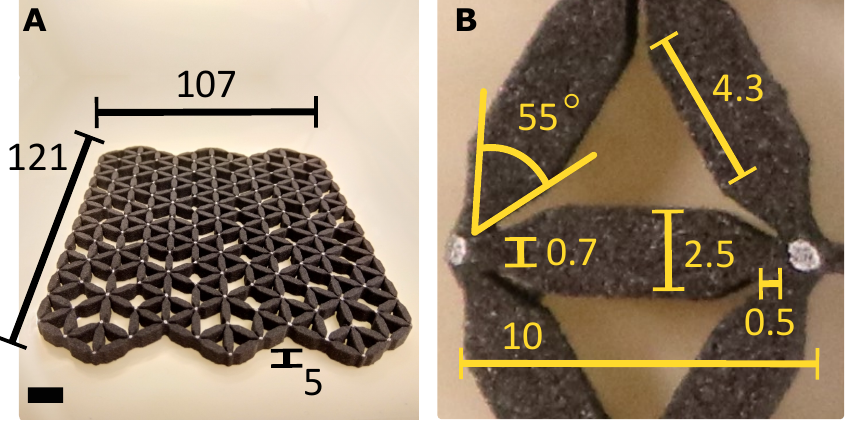}
	\caption{ {\bfseries Experimental network design.} (\textbf{A}) The typical overall dimensions of a 3D printed network are shown in millimeters with an error of $2~\mathsf{mm}$. Scale bar: $10~\mathsf{mm}$. (\textbf{B}) The network's bonds are realized as thick bars tapering to thin hinges that meet at a node (marked in white). Typical dimensions of the bar and connecting hinge are indicated in millimeters with an error of $0.2~\mathsf{mm}$.
	}
	\label{fig_meth_A01}
\end{figure}

\begin{figure}[htb!]
	\centering
	\includegraphics[width=\linewidth]{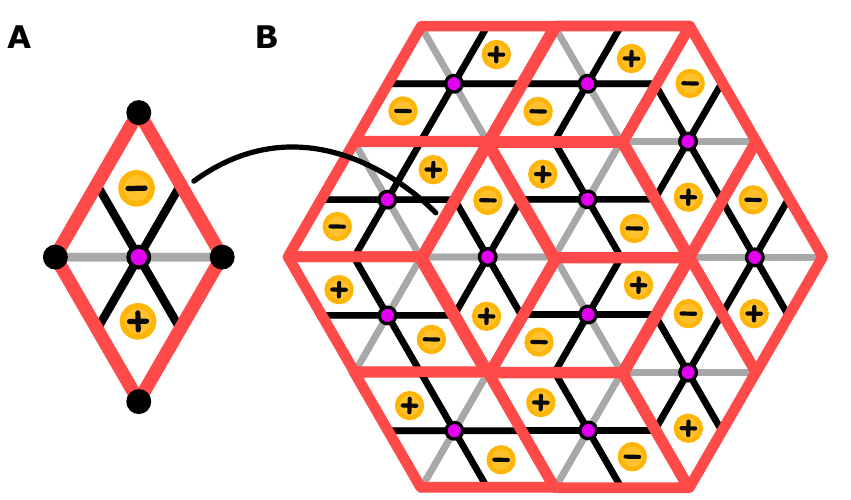}
	\caption{ {\bfseries Counting antiferromagnetic configurations.} (\textbf{A}) All compatible metamaterials with antiferromagnetic spin configurations can be regarded as tilings of diamond-shaped elements (red outline) containing two building blocks. The blocks' minority nodes (pink) are positioned on their shared edge. The blocks deform antiferromagnetically, expanding or contracting (orange markers). (\textbf{B}) Tiling with diamonds produces a hexagonal antiferromagnetic metamaterial.
	}
	\label{fig_af_configs}
\end{figure}

\begin{figure}[htb!]
	\centering
	\includegraphics[width=\linewidth]{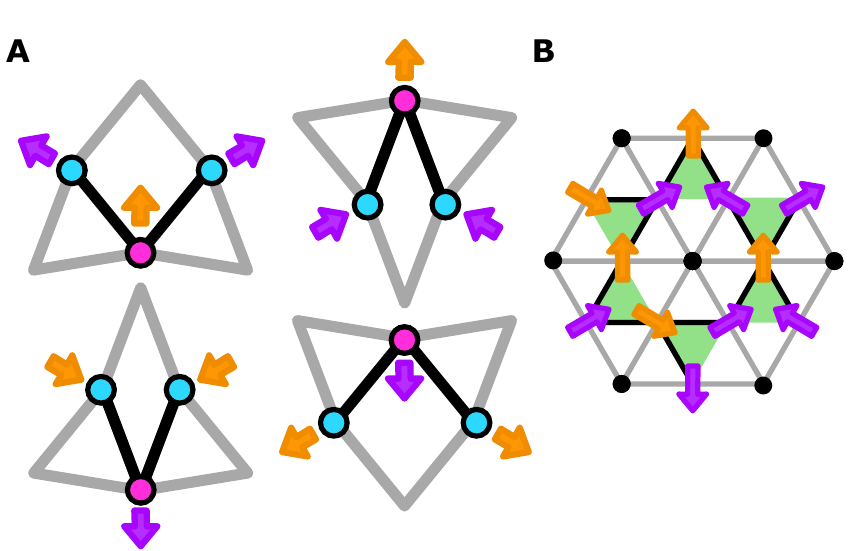}
	\caption{ {\bfseries Antiferromagnetic Ising kagome mapping.} (\textbf{A}) Freely hinging deformations of a building block are represented by edge spins (purple and orange arrows) on its edge nodes (blue and pink). Edges connected by an internal bond have opposite edge spin. For upward-pointing building blocks, orange (purple) denotes inward (outward) deformation, while for downward-pointing building blocks orange (purple) denotes outward (inward) deformation. Shared edge spins of compatibly deformed building blocks are consistent. (\textbf{B})  A compatible architecture and its edge spin state, a ground state of the antiferromagnetic Ising (AFIK) model. The edge spin sites form a kagome lattice (green triangles).
	}
	\label{fig_isingspin}
\end{figure}

\begin{figure}[htb!]
	\centering
	\includegraphics[width=\linewidth]{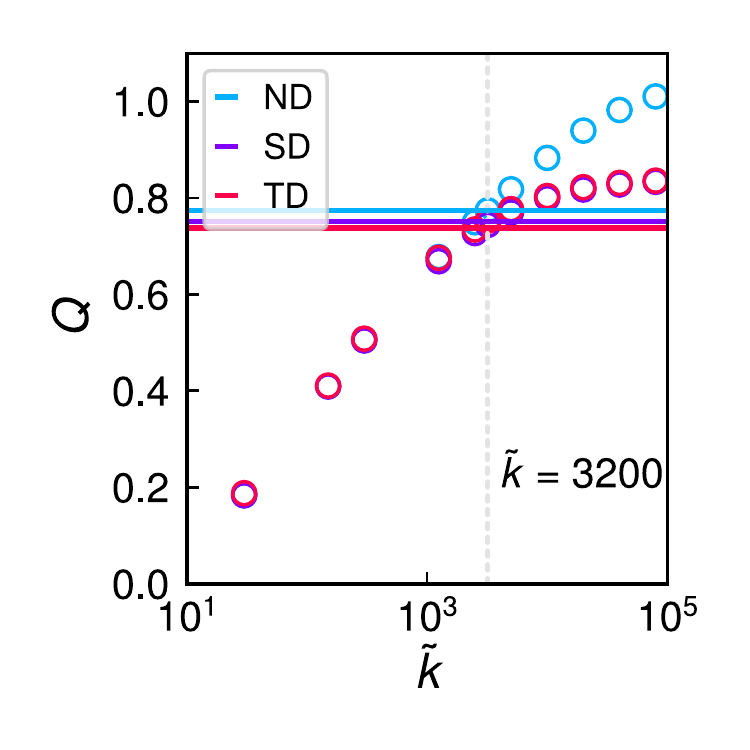}
	\caption{ {\bfseries Stiffness ratio.} The network response predicted by model TR yields a detection factor $Q$ (main text) that varies with the stiffness ratio $\tilde{k}$ between the bonds' axial stiffness and the nodes' hinging stiffness. Comparison of results from model TR (circles) and experiments (solid line) indicate that stiffness ratio $\tilde{k}\approx 3200$ (dashed line) yields the best match between experiment and model for networks with no defect (ND), and a good match for networks with a structural (SD) or topological (TD) defect.
	}
	\label{fig_meth01}
\end{figure}

\begin{figure}[htb!]
	\centering
	\includegraphics[width=\linewidth]{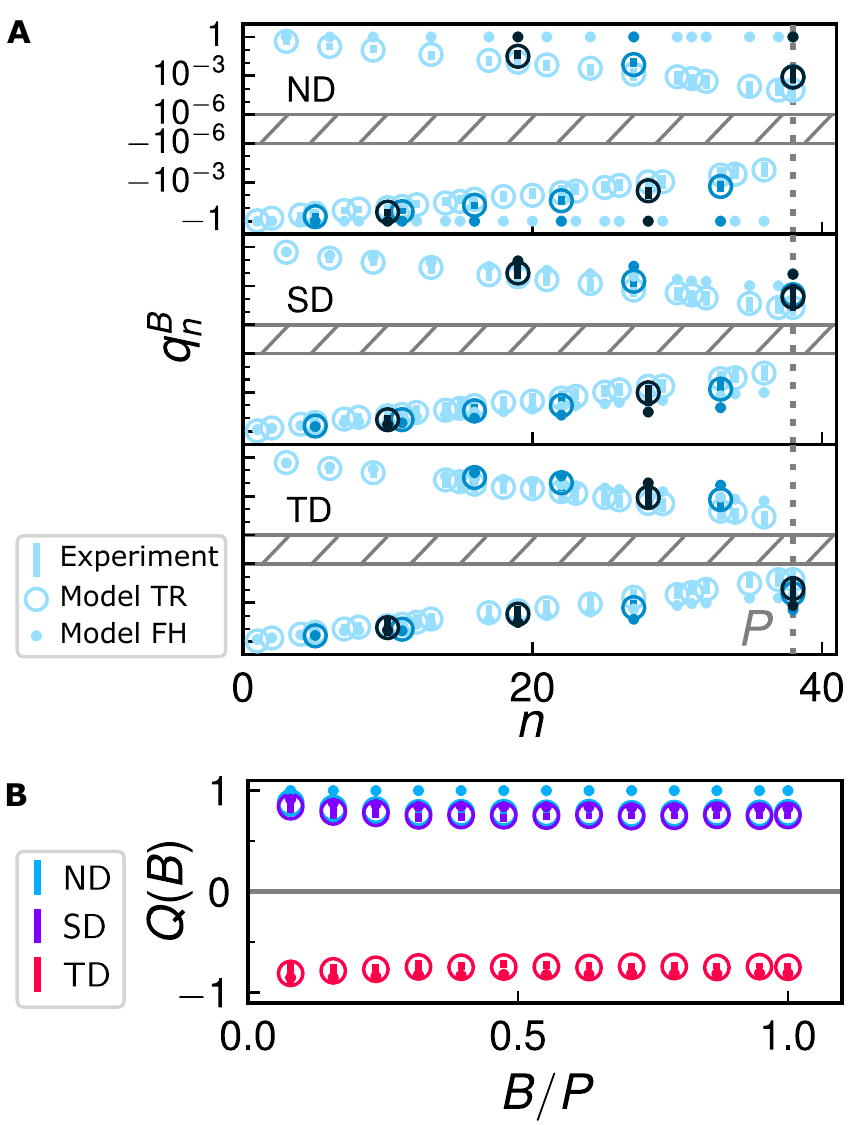}
	\caption{ {\bfseries Topological detection.} (\textbf{A}) The diluted cumulative transfer product $q^B_n$ using the aperiodic designs containing no defect (ND, top), a structural defect (SD, middle) or a topological defect (TD, bottom)
		shown in Fig.~2, A to C. We compare data for $B=4$ (black), $B=7$ (blue) and $B=P=38$ (light blue) edge blocks.
		For the experimental data, a block strain of $\delta = 0.34 \pm 0.04$ was imposed on each probed boundary block; for the calculations using model TR, $\delta=0.23$ and stiffness ratio $\tilde{k}=3200$ were chosen; for the linearized model FH, the results are independent of the block strain to leading order. (\textbf{B}) The topological detection scheme performed on a fraction of $B/P$ of all boundary blocks produces a diluted order parameter $Q(B)$. The sign of the order parameter is negative only when a topological defect is present (legend), while its magnitude is nearly independent of $B/P$.
	}
	\label{fig_diluted}
\end{figure}

\begin{figure*}[htb!]
	\centering
	\includegraphics[width=.99\linewidth]{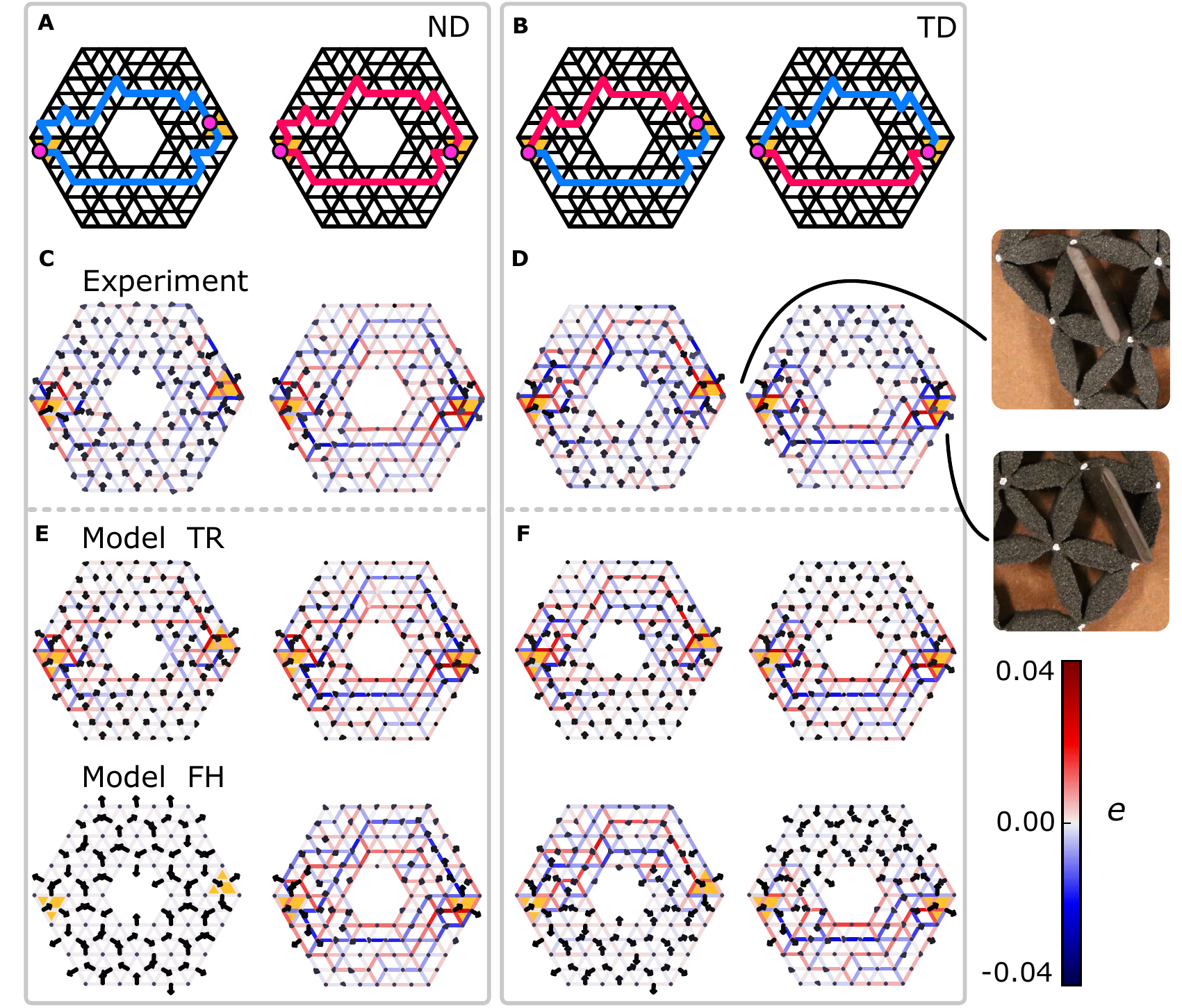} 
	\caption{ {\bfseries Stress steering.} (\textbf{A}) Compatible network design with no defect (ND). A pair of two blocks (orange triangles) undergoes an identical ferromagnetic (blue lines) or antiferromagnetic (red) interaction along both sides of the network. (\textbf{B}) The block pair undergoes an opposite interaction along the two network sides when a topological defect is present (TD). (\textbf{C}) In experimental samples, the block pair is forced ferromagnetically by expanding both to a block strain $\delta = 0.35\pm 0.04$ (inset). Experimentally measured node displacements (arrows, tripled in size for clarity) and relative bond elongations (color bar) under forced expansion of the blocks are shown. The antiferromagnetic interaction paths are frustrated and concentrate stress, while the ferromagnetic paths are unfrustrated and able to deform: the compatible network either deforms globally or is fully stressed. (\textbf{D}) The defected network deforms in one half, while the other half is stressed. (\textbf{E}) (\textbf{F}) Experimental results are reproduced quantitatively in model TR (top row): the elongation and displacement fields' cosine similarities vary from 70 to 98 per cent. Model FH (bottom row), which includes neither hinging stiffness nor concomitant decay, matches experiments qualitatively. Block strains $\delta = 0.35$ were used in both models. Node displacements (arrows, tripled in size) and bond elongations (color bar) are indicated.
	}
	\label{fig_meth_A03}
\end{figure*}


\begin{thebibliography}{10}

	
	\bibitem{Mullin2007}
	T.~Mullin, S.~Deschanel, K.~Bertoldi, M.~C. Boyce, {\it Phys. Rev. Lett\/} {\bf
		99}, 1 (2007).
	
	\bibitem{Shim2012}
	J.~Shim, C.~Perdigou, E.~R. Chen, K.~Bertoldi, P.~M. Reis, {\it Proc. Natl.
		Acad. Sci. USA\/} {\bf 109}, 5978 (2012).
	
	\bibitem{Coulais2016a}
	C.~Coulais, E.~Teomy, K.~{De Reus}, Y.~Shokef, M.~{Van Hecke}, {\it Nature\/}
	{\bf 535}, 529 (2016).
	
	\bibitem{Dudte2016c}
	L.~H. Dudte, E.~Vouga, T.~Tachi, L.~Mahadevan, {\it Nat. Mater.\/} {\bf 15},
	583 (2016).
	
	\bibitem{Chen2016}
	B.~G.-g. Chen, {\it et~al.\/}, {\it Phys. Rev. Lett.\/} {\bf 116}, 135501
	(2016).
	
	\bibitem{Paulose2015a}
	J.~Paulose, B.~G.-g. Chen, V.~Vitelli, {\it Nat. Phys.\/} {\bf 11}, 153 (2015).
	
	\bibitem{Paulose2015}
	J.~Paulose, A.~S. Meeussen, V.~Vitelli, {\it Proc. Natl. Acad. Sci. USA\/} {\bf
		112}, 7639 (2015).
	
	\bibitem{Serra-Garcia2018}
	M.~Serra-Garcia, {\it et~al.\/}, {\it Nature\/} {\bf 555}, 342 (2018).
	
	\bibitem{Coulais2018c}
	C.~Coulais, A.~Sabbadini, F.~Vink, M.~Van~Hecke, {\it Nature\/} {\bf 561}, 512
	(2018).
	
	\bibitem{Florijn2014a}
	B.~Florijn, C.~Coulais, M.~{Van Hecke}, {\it Phys. Rev. Lett.\/} {\bf 113},
	175503 (2014).
	
	\bibitem{Silverberg2014}
	J.~L. Silverberg, {\it et~al.\/}, {\it Science\/} {\bf 345}, 647 (2014).
	
	\bibitem{Frenzel2017}
	T.~Frenzel, M.~Kadic, M.~Wegener, {\it Science\/} {\bf 358}, 1072 (2017).
	
	\bibitem{Bertoldi2017}
	K.~Bertoldi, V.~Vitelli, J.~Christensen, M.~{Van Hecke}, {\it Nat. Rev.
		Mater.\/} {\bf 2}, 17066 (2017).
	
	\bibitem{Kang2014}
	S.~H. Kang, {\it et~al.\/}, {\it Phys. Rev. Lett\/} {\bf 112}, 1 (2014).
	
	\bibitem{Nisoli2013}
	C.~Nisoli, R.~Moessner, P.~Schiffer, {\it Rev. Mod. Phys.\/} {\bf 85}, 1473
	(2013).
	
	\bibitem{Wang2006}
	R.~F. Wang, {\it et~al.\/}, {\it Nature\/} {\bf 439}, 303 (2006).
	
	\bibitem{Castelnovo2008}
	C.~Castelnovo, R.~Moessner, S.~L. Sondhi, {\it Nature\/} {\bf 451}, 42 (2008).
	
	\bibitem{Grima2005}
	J.~N. Grima, A.~Alderson, K.~E. Evans, {\it Phys. Status Solidi B\/} {\bf 242},
	561 (2005).
	
	\bibitem{Coulais2018a}
	C.~Coulais, C.~Kettenis, M.~{Van Hecke}, {\it Nat. Phys.\/} {\bf 14}, 40
	(2018).
	
	\bibitem{Bertoldi2010}
	K.~Bertoldi, P.~M. Reis, S.~Willshaw, T.~Mullin, {\it Adv. Mater.\/} {\bf 22},
	361 (2010).
	
	\bibitem{Syo51}
	I.~Sy{\^{o}}zi, {\it Prog. Theor. Phys.\/} {\bf 6}, 306 (1951).
	
	\bibitem{Kan53}
	K.~Kano, S.~Naya, {\it Prog. Theor. Phys.\/} {\bf 10}, 158 (1953).
	
	\bibitem{Riv16}
	R.~M.~A. Zandbergen, {On the Number of Configurations of Triangular
		Mechanisms}, Thesis, Leiden University (2016).
	
	\bibitem{Mermin1979}
	N.~D. Mermin, {\it Rev. Mod. Phys.\/} {\bf 51}, 591 (1979).
	
	\bibitem{Alexander2012}
	G.~P. Alexander, B.~G.-g. Chen, E.~A. Matsumoto, R.~D. Kamien, {\it Rev. Mod.
		Phys.\/} {\bf 84}, 497 (2012).
	
	\bibitem{Kane2013}
	C.~L. Kane, T.~C. Lubensky, {\it Nat. Phys.\/} {\bf 10}, 39 (2013).
	
	\bibitem{Ning2018}
	X.~Ning, {\it et~al.\/}, {\it Adv. Mater. Interfaces\/} {\bf 5}, 1 (2018).
	
	\bibitem{McEvoy2015}
	M.~A. McEvoy, N.~Correll, {\it Science\/} {\bf 347}, 1261689 (2015).
	
	\bibitem{Reis2015}
	P.~M. Reis, H.~M. Jaeger, M.~Van~Hecke, {\it Extreme Mech. Lett.\/} {\bf 5}, 25
	(2015).
	
	\bibitem{Wehner2016}
	M.~Wehner, {\it et~al.\/}, {\it Nature\/} {\bf 536}, 451 (2016).
	
\end{thebibliography}

\clearpage

\begin{thebibliography}{10}
	
	\expandafter\ifx\csname url\endcsname\relax
	\def\url#1{\texttt{#1}}\fi
	\expandafter\ifx\csname urlprefix\endcsname\relax\def\urlprefix{URL }\fi
	\providecommand{\bibinfo}[2]{#2}
	\providecommand{\eprint}[2][]{\url{#2}}
	
	\addtocounter{enumiv}{30}
	
	\bibitem{Sinterit}
	{Sinterit sp. z o.o.}, {Sinterit LISA product specification} (2014).
	
	\bibitem{Sinterit_flexa}
	{Sinterit sp. z o.o.}, {Sinterit Flexa Black specification} (2014).
	
	\bibitem{Blunt1077}
	M.~O. Blunt, {\it et~al.\/}, {\it Science\/} {\bf 322}, 1077 (2008).
	
	\bibitem{macmahon}
	P.~A. MacMahon, {\it Combinatory Analysis\/}, vol.~2 (Cambridge University
	Press, London, 1916).
	
	\bibitem{oeis}
	N.~Sloane, {The On-Line Encyclopedia of Integer Sequences} (1996).
	{https://oeis.org/A008793}.
	
	\bibitem{Pellegrino1993}
	S.~Pellegrino, {\it Int. J. Solids Struct.\/} {\bf 30}, 3025 (1993).
	
	\bibitem{Audoly2010_bend}
	B.~Audoly, Y.~Pomeau, {\it {Elasticity and Geometry}\/} (Oxford University
	Press, New York, 2010).
	
\end{thebibliography}
\end{document}